\documentclass[twoside,twocolumn,9pt]{article}
\usepackage{dcolumn,float,bm,epsfig,verbatim,color,soul,multicol} %added lauren
\usepackage{extsizes}
\usepackage[super,sort&compress,comma]{natbib} 
\usepackage[version=3]{mhchem}
\usepackage[left=1.5cm, right=1.5cm, top=1.785cm, bottom=2.0cm]{geometry}
\usepackage{balance}
\usepackage{mathptmx}
\usepackage{sectsty}
\usepackage{graphicx} 
\usepackage{lastpage}
\usepackage[format=plain,justification=justified,singlelinecheck=false,font={stretch=1.125,small,sf},labelfont=bf,labelsep=space]{caption}
\usepackage{float}
\usepackage{fancyhdr}
\usepackage{fnpos}
\usepackage[english]{babel}
\addto{\captionsenglish}{ }
\usepackage{array}
\usepackage{droidsans}
\usepackage{charter}
\usepackage[T1]{fontenc}
\usepackage[usenames,dvipsnames]{xcolor}
\usepackage{setspace}
\usepackage[compact]{titlesec}
\usepackage{hyperref}
\usepackage{epstopdf}%This line makes .eps figures into .pdf - please comment out if not required.

\definecolor{cream}{RGB}{222,217,201}

\begin{document}

\pagestyle{fancy}
\thispagestyle{plain}
\fancypagestyle{plain}{
%%%HEADER%%%
\renewcommand{\headrulewidth}{0pt}
}
%%%END OF HEADER%%%

%%%PAGE SETUP - Please do not change any commands within this section%%%
\makeFNbottom
\makeatletter
\renewcommand\LARGE{\@setfontsize\LARGE{15pt}{17}}
\renewcommand\Large{\@setfontsize\Large{12pt}{14}}
\renewcommand\large{\@setfontsize\large{10pt}{12}}
\renewcommand\footnotesize{\@setfontsize\footnotesize{7pt}{10}}
\makeatother

\renewcommand{\thefootnote}{\fnsymbol{footnote}}
\renewcommand\footnoterule{\vspace*{1pt}% 
\color{cream}\hrule width 3.5in height 0.4pt \color{black}\vspace*{5pt}} 
\setcounter{secnumdepth}{5}

\makeatletter 
\renewcommand\@biblabel[1]{#1}            
\renewcommand\@makefntext[1]% 
{\noindent\makebox[0pt][r]{\@thefnmark\,}#1}
\makeatother 
\renewcommand{\figurename}{\small{Fig.}~}
\sectionfont{\sffamily\Large}
\subsectionfont{\normalsize}
\subsubsectionfont{\bf}
\setstretch{1.125} %In particular, please do not alter this line.
\setlength{\skip\footins}{0.8cm}
\setlength{\footnotesep}{0.25cm}
\setlength{\jot}{10pt}
\titlespacing*{\section}{0pt}{4pt}{4pt}
\titlespacing*{\subsection}{0pt}{15pt}{1pt}
%%%END OF PAGE SETUP%%%

%%%FOOTER%%%
\fancyfoot{}
\fancyfoot[LO,RE]{\vspace{-7.1pt}\includegraphics[height=9pt]{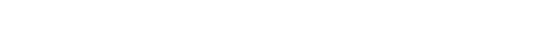}}
\fancyfoot[CO]{\vspace{-7.1pt}\hspace{13.2cm}\includegraphics{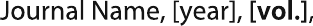}}
\fancyfoot[CE]{\vspace{-7.2pt}\hspace{-14.2cm}\includegraphics{head_foot/RF}}
\fancyfoot[RO]{\footnotesize{\sffamily{1--\pageref{LastPage} ~\textbar  \hspace{2pt}\thepage}}}
\fancyfoot[LE]{\footnotesize{\sffamily{\thepage~\textbar\hspace{3.45cm} 1--\pageref{LastPage}}}}
\fancyhead{}
\renewcommand{\headrulewidth}{0pt} 
\renewcommand{\footrulewidth}{0pt}
\setlength{\arrayrulewidth}{1pt}
\setlength{\columnsep}{6.5mm}
\setlength\bibsep{1pt}
%%%END OF FOOTER%%%

%%%FIGURE SETUP - please do not change any commands within this section%%%
\makeatletter 
\newlength{\figrulesep} 
\setlength{\figrulesep}{0.5\textfloatsep} 

\newcommand{\topfigrule}{\vspace*{-1pt}% 
\noindent{\color{cream}\rule[-\figrulesep]{\columnwidth}{1.5pt}} }

\newcommand{\botfigrule}{\vspace*{-2pt}% 
\noindent{\color{cream}\rule[\figrulesep]{\columnwidth}{1.5pt}} }

\newcommand{\dblfigrule}{\vspace*{-1pt}% 
\noindent{\color{cream}\rule[-\figrulesep]{\textwidth}{1.5pt}} }

\makeatother
%%%END OF FIGURE SETUP%%%

%%%TITLE, AUTHORS AND ABSTRACT%%%
\twocolumn[
  \begin{@twocolumnfalse}
{\includegraphics[height=30pt]{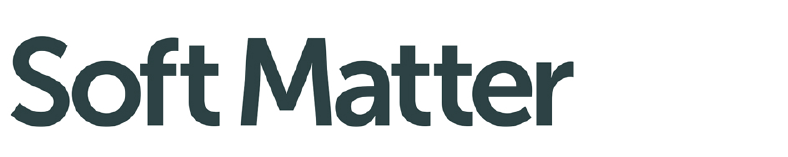}\hfill\raisebox{0pt}[0pt][0pt]{\includegraphics[height=55pt]{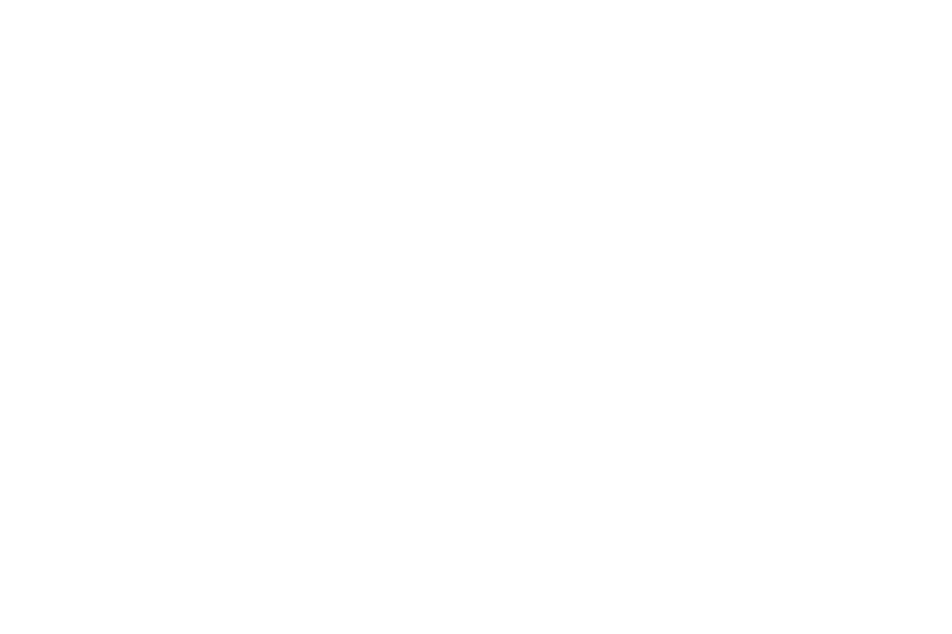}}\\[1ex]
\includegraphics[width=18.5cm]{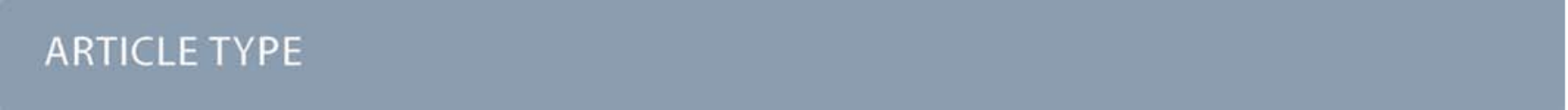}}\par
\vspace{1em}
\sffamily
\begin{tabular}{m{4.5cm} p{13.5cm} }

\includegraphics{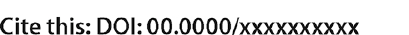} & \noindent\LARGE{\textbf{Sustained Order-Disorder Transitions in a Model Colloidal System Driven by Rhythmic Crosslinking}} \\%Article title goes here instead of the text "This is the title"
\vspace{0.3cm} & \vspace{0.3cm} \\

 & \noindent\large{Lauren Melcher,\textit{$^{b}$} Elisabeth Rennert,$^{\ddag}$\textit{$^{c}$}, Jennifer Ross,\textit{$^{d}$} Michael Rust,\textit{$^{e}$}, Rae Robert-Anderson,\textit{$^{f}$} and Moumita Das \textit{$^{a}$}} \\%Author names go here instead of "Full name", etc.

\includegraphics{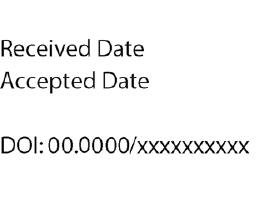} & \noindent\normalsize{Biological systems have the unique ability to self-organize and generate autonomous motion and work. Motivated by this, we investigate a 2D model colloidal network that can repeatedly transition between disordered states of low connectivity and ordered states of high connectivity via rhythmic binding and unbinding of biomimetic crosslinkers. We use Langevin dynamics to investigate the time-dependent changes in structure and collective properties of this system as a function of colloidal packing fractions and crosslinker oscillation periods and characterize the degree of order in the system by using network connectivity, bond length distributions, and collective motion. Our simulations suggest that we can achieve distinct states of this colloidal system with pronounced differences in microstructural order and large residence times in the ordered state when crosslinker kinetics and lifetimes depend directly on the oscillation period and this oscillation period is much larger than the colloidal diffusion time. Our results will provide insights into
the rational design of smart active materials that can independently cycle between ordered and disordered states with desired material properties on a programmed schedule.} \\%The abstrast goes here instead of the text "The abstract should be..."

\end{tabular}

 \end{@twocolumnfalse} \vspace{0.6cm}

  ]
%%%END OF TITLE, AUTHORS AND ABSTRACT%%%

%%%FONT SETUP - please do not change any commands within this section
\renewcommand*\rmdefault{bch}\normalfont\upshape
\rmfamily
\section*{}
\vspace{-1cm}

%%%FOOTNOTES%%%

\footnotetext{\textit{$^{a}$~ School of Physics and Astronomy, Rochester Institute of Technology, Rochester, NY, USA; E-mail: modsps@rit.edu}}
\footnotetext{\textit{$^{b}$~ School of Mathematical Sciences, Rochester Institute of Technology, Rochester, NY, USA}}
\footnotetext{\textit{$^{c}$~ School of Chemical Sciences, and School of Physics and Astronomy, Rochester Institute of Technology, Rochester, NY, USA}}
\footnotetext{\textit{$^{d}$~  Department of Physics, Syracuse University, Syracuse, NY, USA}}
\footnotetext{\textit{$^{e}$~  Department of Biology, University of Chicago, Chicago, IL,  USA}}
\footnotetext{\textit{$^{d}$~  Department of Physics and Biophysics, University of San Diego, San Diego, CA, USA}}
\footnotetext{\dag~Electronic Supplementary Information (ESI) available: [details of any supplementary information available should be included here]. See DOI: 10.1039/cXsm00000x/}
\footnotetext{\ddag~Present address: University of Chicago, Chicago, IL, USA}

%%%END OF FOOTNOTES%%%

%%%MAIN TEXT%%%%
% The main text of the article\cite{Mena2000} should appear here.
\section{Background}
Living organisms have active components that allow for generation of coherent motion and self-organized dynamics.\cite{alvarado2013molecular} A mechanistic understanding of these biological materials can advance the engineering of new smart materials that can self-organize, self-stimulate, and evolve independently according to the application environment. In the past decade, studies of active soft materials have sought to understand the layers of complexities and mechanistic structures in biological systems.\cite{needleman2017active,buttinoni2013dynamical,bialke2015active} However, the rational design of active synthetic materials that can recapitulate the autonomous, persistent, and reversible dynamics of living materials without external intervention remains a challenge. 

\par Colloidal systems offer an attractive platform for constructing bottom-up minimal models that can offer insights into the rational design of such materials, given colloidal suspension and self-assembly have been studied extensively and are well understood.\cite{marcus1996observations,chen2011directed} Most studies of colloidal systems, however, investigate phase separations and transitions that were triggered by external actuation via changes in temperature, light, and pressure.\cite{ebbens2016active,zaccarelli2006gel}.  Furthermore, most active material prototypes and models are unable to attain reversible transitions between states at a level consistent with living systems. For example, while some experiments and models of colloids do incorporate polymer attachments, the exceedingly small timescales ($\sim$ milliseconds to seconds) of these polymeric attachments could not induce the reversibility of states \cite{berner2018oscillating,Abernethy2003} which is a hallmark of living systems. 

\par What are the principles underlying the rational design of a system that can repeatedly transition between states with distinct structural and material properties at reliable user-defined periods, without the need of external actuators? To obtain insights into this question, we construct and investigate a model system consisting of colloids that can be connected via crosslinkers with oscillatory dynamics such that they can autonomously cycle between states with distinct degree of order and material properties. 
The crosslinkers regulating these oscillatory dynamics in our computational model are inspired by the bacterial clock proteins, KaiABC. These circadian oscillator proteins are found in cyanobacteria \textit{S. elongatus} and have demonstrated a steady, regular 24-hr rhythm in vitro.\cite{chen2015transplantability} Here, we demonstrate the potential for effectively harnessing such rhythms to drive transitions in material properties of model colloidal systems.
%to achieve materials with robust time-dependent oscillations. 
Through our predictive mathematical modeling of persistent order-disorder transitions in such systems of colloids and rhythmic crosslinkers, we hope to inform experiments by predicting the time-evolving collective properties as a function of crosslinker oscillation periods and colloidal packing fractions. Our results will also provide insights into the mechanisms that lead to emergent properties in these systems.

\section{Model and Method}

\subsection{Mathematical Model}

We simulate a model two-dimensional system consisting of colloidal particles and crosslinkers with oscillatory kinetics. The main ingredients and assumptions are as follows:
\begin{enumerate} 
\item The colloidal particles interact with each other via Lennard Jones (LJ) interaction potential $V_{LJ}$ \cite{wang2020lennard}, given by 
 \begin{equation}
  {V}_{LJ}= 4\epsilon\left[ \left (\frac{\sigma}{r_{ij}}\right)^{12} -\left(\frac{\sigma}{r_{ij}}\right)^{6}\right],
\end{equation}

where $r_{ij}$ is the interparticle distance, $\sigma$ is the particle diameter, and $\epsilon$ is the depth of the potential well. To reduce the computational expense, for our model, the interparticle distance $r_{ij}$ is cut off at $2^{\frac{1}{6}}$, which encodes the resulting force as a repulsive-only force. 

\item When two particles are within a distance less than or equal to the rest length $r_{0}$ of the crosslinkers, they can become connected by a crosslinker with an attachment probability $P_{a}$. The probability is dependent on the fraction of free crosslinkers $p_{k}$  available to connect the particles and the time-period $T$ of the crosslinkers as $P_{a}=p_{k}p_{0}\sin^{2}(\pi t/T)$, where 
$p_{0}$ is a phenomenological parameter corresponding to the initial crosslinker concentration. We assume the crosslinker is dilute on the particle surface; while there are multiple crosslinking attachment sites on each colloid, we limit the number of crosslinks between any two particles to one.

\item Once two particles are crosslinked, they stay attached for a given lifetime, unless the bonds are stretched beyond a length of 3$\sigma$, leading to particles unlinking prematurely. For simplicity, we associate the timescale of oscillations to also govern the lifetime of the crosslinkers. This lifetime is drawn from a Gaussian Distribution with a mean of $T/2$ and a standard deviation of $T/8$, where $T$ is the crosslinker oscillation period mentioned earlier. This ensures the particles remain attached for a significant amount of time and avoids too many particles simultaneously attaching or detaching. 

\item The particle motion follows the overdamped Langevin equation:
\begin{equation}
    \frac{d\textbf{r}}{dt}=\frac{D}{K_{B}T_{R}} (\textbf{F}_{LJ}+\textbf{F}_{c})+\sqrt{2D}\eta, 
    \label{EoM}
\end{equation}
where the interparticle interaction force consists of two components: a force $\textbf{F}_{LJ}$ obtained from the Lennard Jones potential defined earlier, and an elastic force $\textbf{F}_{c}$ due to the stretching or compression of the crosslinker connecting the particles which is modeled as a Hookean spring\cite{ma2018structural}. The latter is defined as $\textbf{F}_{c}=-K(r_{ij}-r_{0})$, where 
 $r_{0}$ is the crosslinker rest length and $K$ is the spring coefficient. The constants $K_{B}$ and $T_{R}$ represent the Boltzmann constant and room temperature, respectively. The diffusive motion of the particles is described by the third term $\sqrt{2D}\eta$, where $D$ is the diffusion coefficient and $\eta$  represents Gaussian noise with zero mean and unit variance.  
%  In our simulations, the length of the bond connecting two colloids is the sum of the radii of the two colloids and the length of the biophysical crosslinker between them; for colloids of radius 1 $\mu m$, this bond length is approximately 1.01 $\mu m$, assuming a length of 10$nm$ which is the size of typical crosslinking proteins. We set the value of the spring constant
 
%  The crosslinker rest length is 1.01 since the bond length is the distance between the centers of a pair of crosslinked colloids and the length of the crosslinker connecting them is ~$10n$m. 
 
%  We set $K$ equal to 9 so the forces arising from the LJ interactions and spring forces are of comparable magnitude. The diffusive motion of the particles is described by the third term $\sqrt{2D}\eta$, where $D$ is the diffusion coefficient and $\eta$  represents Gaussian noise with zero mean and unit variance.  

\end{enumerate}
 
\begin{figure}
    \centering
    \includegraphics[width=\columnwidth]{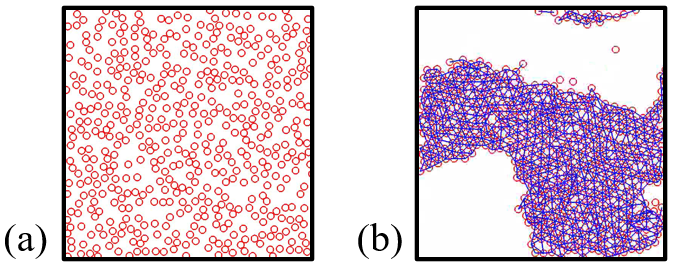}
      \caption{ A model schematic of the colloidal system with a packing fraction of 30\% and $T=18000$. (a) The colloids are initially unlinked and diffusing in the simulation box. (b) When the system reaches an ordered state, the colloids are crosslinked via the oscillating crosslinkers}
    \label{modelSchematic}
\end{figure}

We initialize the system with five hundred unlinked colloid particles randomly distributed in the simulation box. When two colloidal particles are crosslinked, the length of the bond connecting them is the sum of the radii of the colloids and the length of the biomimetic crosslinker between them; for colloids of radius $1\mu m$, this bond length is approximately $1.01\mu m$, assuming a crosslinker length of $10nm$ which is the size of typical proteins. We set the value of the spring constant of this bond to be $0.036pN/\mu m$ such that the forces arising from the LJ interactions and spring forces are of comparable magnitude. We solve equation \ref{EoM} for the particle positions, using first-order Euler integration. Period boundary conditions are used to avoid the computational expense of a large number of particles or a large simulation box size.\cite{wu2014applying}. The ratio of the number of colloids to crosslinkers is set to $1:4$ to allow for a considerable amount of crosslinking of the colloids while also ensuring the system is not over-saturated with crosslinkers and can revert back to an unlinked, disordered state.  This attachment probability $P_{a}$ is calculated at every time step across all the particles within $r_{0}$ of each other, and the phenomenological parameter $p_{0}$ is set to $0.005$  to yield clear transitions between crosslinked and unlinked states, as illustrated in Fig. \ref{modelSchematic}. The diffusion constant is set to $0.483\mu m^{2}/s$ to be within experimentally meaningful range.  All of the lengths are scaled by the diameter of a colloid, which is $\sigma = 1\mu m$,  all times by the time $\tau_{D}$ taken by a colloid to freely diffuse across its diameter, which is $2.07$ seconds in our simulations, and energies by $K_{B}T_{R}\sim4pN nm$, which is the energy associated with thermal fluctuations at room temperature. Coming to the period of oscillation $T$ of the crosslinkers, while the duration of a 24-hr circadian rhythm of KaiABC oscillators is equivalent to $43200\tau_{D}$, biological oscillators in nature have a wide range of oscillation periods, including values much smaller than that of the KaiABC proteins \cite{goldbeter2002computational}. To decrease the computational expense, we explore shorter crosslinker oscillation periods, ranging from $T$=$18$ to $T$=$18000$. The main variables and parameters in this model are presented in the following tables below and chosen to mimic a system of $1\mu m$ beads suspended in water with 2D area fractions of $10-30\%$. 

% \begin{table}[h]
% \centering
% \small
%   \label{tbl:example1}
%   \begin{tabular*}{0.5\textwidth}{@{\extracolsep{\fill}}lll}
%     \hline
%     Variable/Parameter & Description & Units \\
%     \hline
% $r_{ij}$ & interparticle distance & $\mu$m\\
% $r_{0}$ & rest length of crosslinker & $\mu$m\\ 
% $K$ & spring coefficient of crosslinker & N$/ \mu$m \\
% $D$ & diffusion coefficient & $\mu$m$^{2}/s$ \\
% $\eta$ & gaussian noise vector \\
% $\beta$ & $\frac{1}{(k_{B}T)}$ ($\frac{1}{energy}$) & $\frac{1}{J}$ \\
% $\epsilon$ & depth of potential well & J \\
% $\sigma$ & particle diameter & $\mu$m\\
% $t$ & unit of time & s \\
%     \hline
%   \end{tabular*}
% \end{table}

\subsection{Varied Quantities}
The two parameters varied in our simulations are the crosslinker oscillation period, and the colloidal packing fraction i.e., the fraction of area occupied by the particles in the simulation box. 
\begin{itemize} 
\item Crosslinker oscillation period: We investigate a model system with a colloid packing fraction fixed at 30\% and $4$ crosslinker oscillation periods $T=18$, $180$, $1800$, and $18000$, which are equivalent to $0.01$, $0.1$, $1$, and $10$ hours, respectively. The colloid concentration is kept intentionally dilute to allow the crosslinker proteins to induce system-wide oscillations between ordered and disordered states; too high of a packing fraction would automatically lead to an ordered state being formed, regardless of crosslinker dynamics.

\item  Packing fraction of colloidal particles: We keep the crosslinker oscillation period fixed at $1800$ while varying the colloid packing fraction from $10-30\%$.
\end{itemize}

\subsection{Measured Quantities}
To determine transitions between liquid-like, disordered states and  gel-like, ordered states, we investigate and quantitatively characterize the degree of order of our model system
 by calculating the connectivity, the ensemble average squared displacement, the mean squared displacement, the radial distribution function, and the bond length distribution as described below.
 
 \begin{itemize} 
 \item  The connectivity is calculated by finding the number of neighbors each colloidal particle is crosslinked to and averaging over all the particles. This average is calculated at every time step and plotted over time. The connectivity informs us of how network-like or connected our colloidal system is throughout the simulation. Large, pronounced amplitudes demonstrate that the system is transitioning between states with significantly different degree of order. 
 
 \item  The ensemble average squared displacement (EASD) measures the network movement as a function of time. The equation for the ensemble average squared displacement is $\big \langle \left[ r(t_{0}+t)-r(t_{0})\right]^{2} \big \rangle$ where $r$ represents the particle position and $t$ represents the time-period of interest. In our model, $t_{0}$ is set to $t=0$ to demonstrate how the system changes over real time, starting from the beginning of the simulation. At each time $t$, we sum the square of the displacements of each particle and then average over the total number of particles.

\item The mean squared displacement (MSD) is a sliding average given by  $\big \langle \left[ r(t+\tau)-r(t)\right]^{2} \big \rangle$ and describes the motion of particles over certain lag times. The particle position at time $t$ is denoted by $r(t)$ and the lag time is given by $\tau$. From this measurement, we measure particle motion over small and large time intervals.

\item The radial distribution function (RDF) describes the probability of finding a particle within a shell of width $dr$ at a distance $r_{n}$ from a reference particle $n$. It is obtained by calculating the distance between all particle pairs and binning them into a histogram. The histogram is normalized with respect to a system with no correlations between particle positions, i.e., where the particles do not interact with each other. From this metric, we gain a better understanding of how ordered our system is. Sharp regular peaks in the plotted results indicate a ordered,  network-like state, while fewer, less distinct peaks mean the system is less-ordered. 

\item The distribution of the lengths of the bonds between particles in the ordered states provides insights into network and cluster properties. A sharply peaked distribution suggests a well-connected network with a high degree of order, while a broad distribution indicates the presence of smaller clusters.
\end{itemize}

We also examine the general particle positions and movement and generate system configurations that illustrate the behavior of the network and provide insight into the structure of the network and clustering for different parameters.

\section{Results and Discussion}

\subsection{Impact of varying crosslinker oscillation period} 

\begin{figure*}[t]
    \centering
    \includegraphics[width=2\columnwidth]{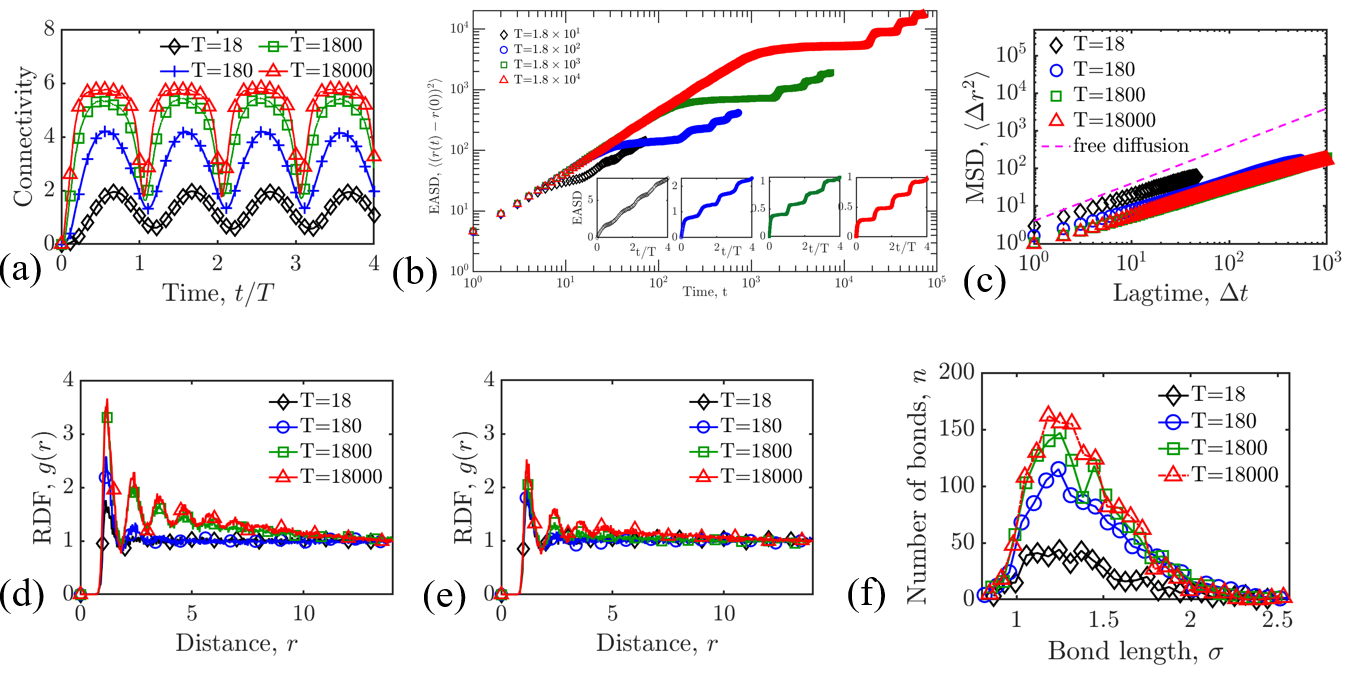}
    \caption{Four systems with different crosslinker oscillation periods ranging from $T=18$ to $18000$ with a packing fraction of $30\%$ are simulated. (a) The levels of connectivity are plotted over time for each of the oscillation periods, measuring the average number of bonds each colloid is connected to over $4$ cycles. (b) The EASD curves are plotted, measuring the network movement beginning at $t=0$. (c) The MSD for the systems, with the free diffusion slope. (d) and (e) The RDF curves are sampled at the ordered and disordered states, respectively. (f) The bond length distributions are sampled at the ordered state.}
    \label{VaryPeriod}
\end{figure*}

First, we discuss the time evolution of the structure and dynamics of the system as we vary the crosslinker oscillation period for a fixed colloid packing fraction of $30\%$.  Our measurements of system-wide connectivity, ensemble average squared displacement, mean squared displacement, radial distribution function, and bond length distribution, shown in Fig. \ref{VaryPeriod}, demonstrate that the system repeatedly transitions between a highly crosslinked, ordered state, and an unlinked, disordered state. 

\par We observe clear oscillations in the average connectivity of the colloidal system, and these oscillations between states of high connectivity and low connectivity are more striking as the oscillation period $T$ is increased as shown in Fig. \ref{VaryPeriod}(a). For example, when $T=18$, the connectivity oscillates between $0.5$ to $2$ indicating that the system stays in a disordered, fluid-like state, and the colloids do not form any noticeable spanning networks. Because the smaller oscillation periods yield shorter bond lifetimes, they effectively reduce the time the system takes to attain peak connectivity. Increasing the crosslinker period to $T=18000$ yields a peak connectivity of $\sim 6$, which corresponds to the largest coordination number for a two-dimensional network in equilibrium, indicating a very ordered state, and a trough of $\sim 2$ indicating a disordered state. Intermediate values of $T$ lead to intermediate values of the peak connectivity and the amplitude of oscillation. We also see that for  $T=180$ the system spends comparable times in the ordered and ordered states, and as we increase $T$, the residence time in the ordered state increases while that in the disordered state decreases. These observations can be attributed to the direct dependence of the lifetime of individual crosslinkers on the crosslinker oscillation period $T$, and to the separation between $T$ and the diffusion time of the colloids. As $T$ increases, both these quantities increase allowing the system to form states with larger and larger connectivities until a maximum value of $6$ is reached, and these connected, ordered states also stay intact for longer times before the crosslinkers come apart and the colloids are unlinked. Since the residence times in an ordered and a disordered state for a given $T$ should add up to give half the oscillation period, as the former increases, the latter decreases.  We also observe that the residual order at the disordered states increases in proportion to the oscillation period. This is a consequence of the residence time imbalance. Larger oscillation periods produce well-ordered states with high levels of connectivity, and since those systems spend severely reduced times in the disordered state, they do not have enough time to dissolve back to a completely unconnected state before the next oscillation cycle starts and the crosslinkers begin to connect particles together.

\begin{figure}[h]
  \centering
  \includegraphics[width=1\columnwidth]{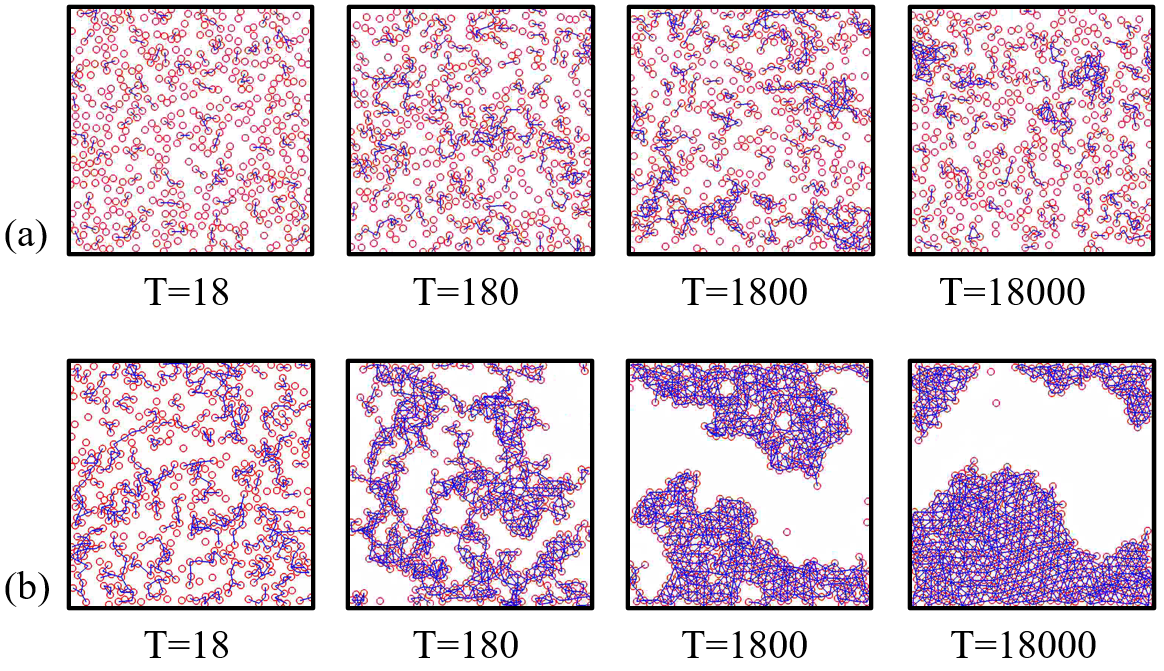}
 \caption{System configurations sampled over (a) the disordered states and (b) the ordered states.}
\label{SysCon_varyT}
\end{figure}
\par When $T=1800$, the fixed lifetime lasts long enough for a spanning cluster formation to occur and although the relaxation time for the colloids is still reduced, the system connectivity drops from nearly $5.5$ to $1.75$ bonds per particle. For the $T=18000$ case, the fixed lifetime of the ordered state is much longer, as demonstrated by the flatter peaks in Fig. \ref{VaryPeriod}(a). While the difference in ordered-disordered state is about $4$ bonds per particle, the relaxation intervals become even shorter, implying that the growing separation between the relaxation timescale and oscillation timescale leads to a greater imbalance of residence times in the ordered and disordered states.

\par We show the ensemble average squared displacement(EASD) as a function of time in Fig. \ref{VaryPeriod}(b). For a given time interval, the slope of the EASD curves indicates the mobility of the system. For the system with the shortest period of $T=18$, the connectivity remains small even in the ordered state, and the difference in the slopes during the ``ordered'' and ``disordered'' intervals is small.  The EASD curves for the systems with the three larger $T$ values, however, show distinct oscillations in mobility, with the plateaus indicating network formation.  For $T=1800$ and $18000$, the slopes of the EASD curves stay $\sim 0$ for longer times than that of $T=180$, which is expected since the smaller period of $180$ corresponds to shorter bond lifetimes, leading to particles having the opportunity to unlink and move about more frequently. The insets in Fig. \ref{VaryPeriod}(b) show the same data but are plotted in linear scale and with the time scaled by the crosslinker oscillation period. This illustrates the patterns of mobility in the system as each oscillation period occurs; here, a stair-stepping pattern can be clearly seen for all the above cases. When the slopes become approximately zero, this signifies the systems' movement slowing down as they cycle up to an ordered state. Then, as the slopes begin to increase, the systems dissolve back to a disordered state. 

\par The MSD curves in Fig. \ref{VaryPeriod}(c) show that the smallest oscillation period $T=18$ yields the highest rate of diffusive motion. This is expected as shorter crosslinker periods produce correspondingly shorter bond lifetimes, so the system is weakly connected throughout the simulation. As $T$  increases, the data shows that the particles are diffusing significantly slower. Because the bond lifetimes are related to the oscillation period, more particles remain connected for longer time intervals and cannot diffuse as freely, consequently resulting in lower rates of diffusive motion. Across all four systems, the particle motion, while still largely diffusive, has a noticeable departure from free diffusion, particularly as the crosslinker time-period increases. 

\begin{figure*}[t]
  \centering
  \includegraphics[width=2.0\columnwidth]{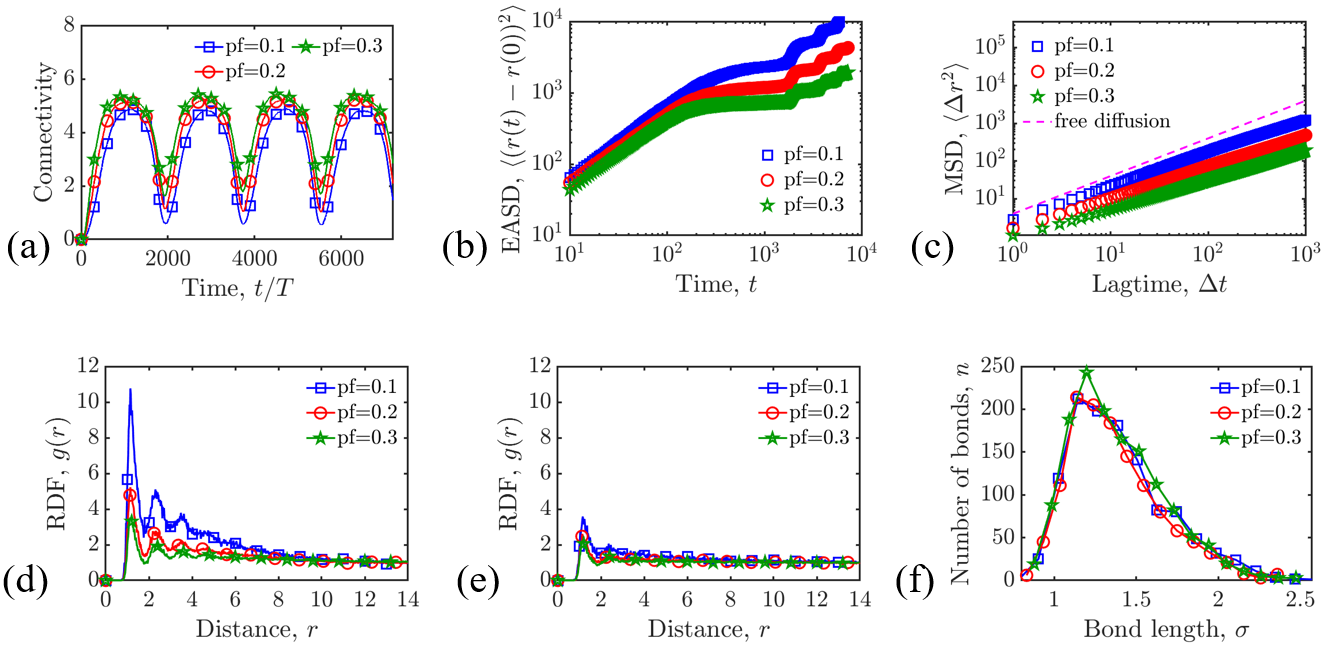}
 \caption{Three systems with different packing fractions ranging from $10\%$ to $30\%$ and a fixed oscillation period of $1800$ are simulated. (a) Connectivity is plotted over time for $3$ packing fractions with the oscillation period $T=1800$. (b) The EASD curves, starting from the beginning of the simulation. (c) The MSD curves, with the free diffusion curve plotted alongside. The RDF is sampled at (d) the ordered state, located at the peaks of the connectivity curve, and (e) the disordered state, occurring at the troughs of the connectivity results. (f) The bond length distribution corresponds to the particle bonds that are formed in the ordered state.}
\label{variedpf}
\end{figure*}

\par To gain more insight into the translational order in the system in the ordered and disordered states, we calculate the RDF for each of the systems. In both Fig. \ref{VaryPeriod}(d) and (e), as the oscillation period increases, the number of peaks increases, signifying the longer oscillation periods yield systems that achieve distinctly well-ordered states, confirming the observations made from the connectivity curves. In the RDF curves, there is a pronounced difference between the ordered and disordered states for $T=1800$ and $18000$. The number of peaks in the curves and the heights and sharpness of those peaks increase in comparison to those in Fig. \ref{VaryPeriod}(e). For the two shorter oscillation periods $T=18$ and $180$, the RDF curves are nearly identical in terms of the peak heights and noise, which implies that the degree of order in the system does not change much between the ordered and disordered states. The colloids do not have the chance to form a large, well-connected network, and stay in a weakly connected state, leading to a subtler distinction between the two states.

\par Next, we study the bond length distributions in the ordered states to understand the connectivity properties of the clusters (see Fig. \ref{VaryPeriod}(f)). For larger $T$, the bond length distributions are narrower. The systems with longer oscillation periods form large single spanning networks in the ordered states, subsequently leading to most particles with $\sim 5 - 6$ nearest neighbors. Because a majority of the particles have a similar high nearest neighbor count, the minimum energy configuration of the system will correspond to a narrow distribution of bonds lengths which is peaked very close to the crosslinker rest length. As the oscillation periods decrease, the bond length distribution becomes wider. This is due to the formation of smaller and less connected clusters at smaller $T$. Since the colloids do not form one large single spanning cluster, some of the particles may have close to $6$ nearest neighbors or connections, while particles belonging to other clusters may have significantly fewer nearest neighbors. 

\par In Fig. \ref{SysCon_varyT}, we show particle configurations at the disordered and ordered states of each of the systems. For the shortest oscillation period $T=18$, the system only transitions between disordered to slightly connected states. In the ordered state, the system has slightly larger connectivity, but the particles do not form distinct clusters. However, as we increase $T$ to $180$ and $1800$, the system begins to oscillate between disordered states with minimal connectivity to ordered states with larger cluster formations. At the longest period $T=18000$, the system can repeatedly form a large, strongly connected network and transition back to a disordered, weakly connected network.

\subsection{Impact of varying colloidal packing fraction}
%need to do verb tense check starting here-lauren
\par We next examine the effect of varying the packing fraction. We explore systems with an oscillation period of $T=1800$ and simulate $3$ cases with a packing fraction range of $10-30\%$. We keep the oscillation period constant at $1800$ so as to allow clearer transitions between ordered and disordered states compared to $T=18$ and $180$ and to reduce the gap between residence times in these states compared to that obtained with $T=18000$. The system is again kept dilute to allow oscillations based on the rhythm of the crosslinkers rather than spontaneous formation of a large connected network, which happens in a dense system.

 \par Higher packing fractions generate systems with more particles within a given distance corresponding to the crosslinker rest length, thereby leading to a larger level of connectivity on average. All three systems in  Fig. \ref{variedpf}(a) can transition between states with $\sim$ $5.5$ to $1$ bonds per particle, which confirms the colloidal networks are able to oscillate between well-connected to very weakly connected states. Across the packing fractions, the systems still produce reduced residence times in the disordered states and comparatively larger durations in the ordered states. The packing fraction does not affect this imbalance of residence times. The results from the ensemble averaging in Fig. \ref{variedpf}(b) show that the EASD steadily increases with time at the beginning of the simulation but after $t=180$, there is decreased mobility in the network as the slopes of the EASD curves become nearly zero, suggesting that particles become locked into a network or cluster relatively early in the first cycle. A stair-stepping phenomenon is present in all $3$ cases, emphasizing the network movement increasing and decreasing over each of the oscillation cycles as the system transitions between ordered to disordered states. The MSD curves show that all three systems are diffusive, but the diffusivity decreases as the packing fraction increases, which is seen clearly when compared against free diffusion in Fig. \ref{variedpf}(c). Systems with the two higher packing fractions have smaller diffusivities since larger colloidal packing fractions lead to larger spanning networks and consequently, more particles have restricted mobility.

 \par We next examine the RDF for the systems at low and high periods of connectivity. There is a considerable difference in the degree of order between the states. As a general trend, as the packing fraction increases, the height of peaks in the RDF curves decreases as a result of a larger normalization factor in the denominator. This effect is seen in both the ordered and disordered states (Fig. \ref{variedpf}(d) and (e)). For the higher packing fractions, large spanning clusters are formed while for lower packing fractions, we observe several smaller clusters, which is why the RDF curve for the $10\%$ packing fraction case has the highest peak at an interparticle distance of $1$. In the disordered state (Fig. \ref{variedpf}(e)), the RDF curves for all three systems have about two discernible peaks and a similar amount of noise, indicating the systems have a low degree of system order during this state. However, in the ordered state, the RDF curves demonstrate the systems achieve a significant increase in the level of order as the number of peaks increases for each of the cases and the sharpness of those peaks is more apparent. The bond length distribution for the ordered states of the system is shown in Fig. \ref{VaryPeriod}(f). The highest packing fraction of $30\%$ yields a well-connected state with a narrow bond length distribution, and the two systems with the smaller packing fractions produce several smaller clusters that are less connected and lead to a wider distribution of bond lengths.
 \begin{figure}[h]
 \centering
 \includegraphics[width=1\columnwidth]{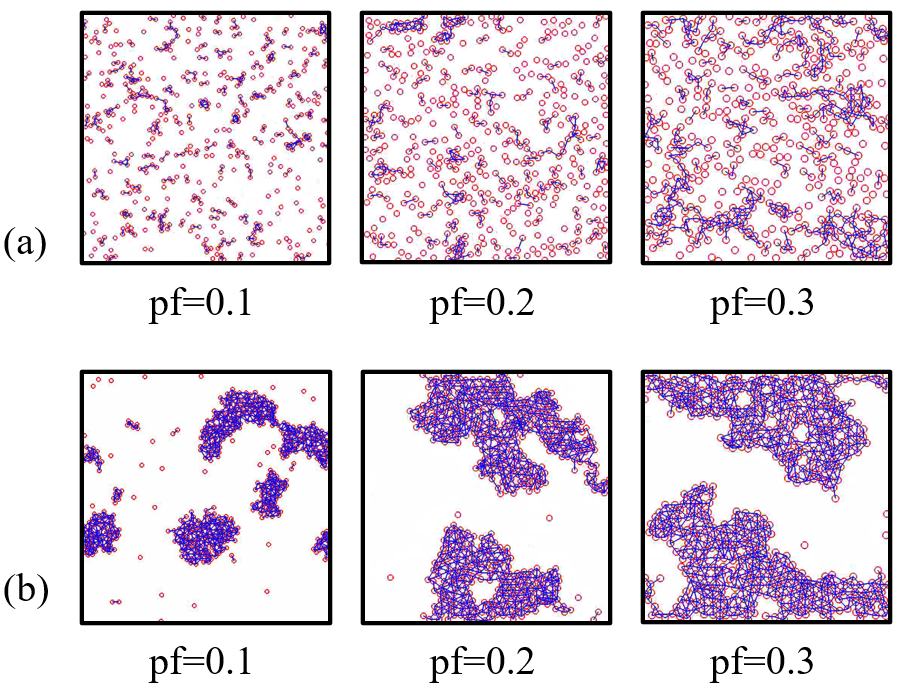}
\caption{System configurations sampled over (a) the disordered states and (b) the ordered states.}
 \label{sysConfigvaryPF}
 \end{figure}
%  \begin{figure}[h]
%  \centering
%  \includegraphics[width=1\columnwidth]{UpdatedFigures/MSD_4parts.png}
% \caption{The MSD is measured over 4 parts: the first half of the cycle, when particles attach and approach a ordered state, the peak, which is when the system is at its most ordered, the second half of the cycle, when particles began to detach and reach a softer state, and the trough, which is the relaxation period of the colloids.}
% \label{MSD_4Parts}
% \end{figure}
\par Sample configurations of each system at low and high  network connectivity are generated for all $3$ systems in Fig. \ref{sysConfigvaryPF}.
The configurations illustrate that the systems tend to oscillate between weakly connected structures with minimal cluster formation and tightly connected networks with bigger clusters. As the packing fraction is increased, the clusters become larger and the particles experience more instances of crosslinking. The size of the clusters in the ordered states is proportional to the packing fraction.

\subsection{Phase Diagrams}
%\subsection{System pf=30\%, T=1800}
% \par The system with a packing fraction of 30\% and an oscillation period of 1800 is investigated further. Although several measurements of this system have been explored already, we decide to analyze the particle motion over different segments of the periodic cycles. In Fig. \ref{MSD_4Parts}, all 4 MSD trajectories are plotted along with the free diffusion curve. The highest rate of diffusion belongs to the trough interval, which is anticipated since a significant amount of particles had unlinked and could diffuse faster as a consequence. The attachment and detachment periods have similar diffusion rates for shorter lag times while the peak interval yields the slowest rate of diffusion when the system reaches a higher level of order and a spanning network had formed.

\par
\begin{figure}[h]
  \centering
  \includegraphics[width=1\columnwidth]{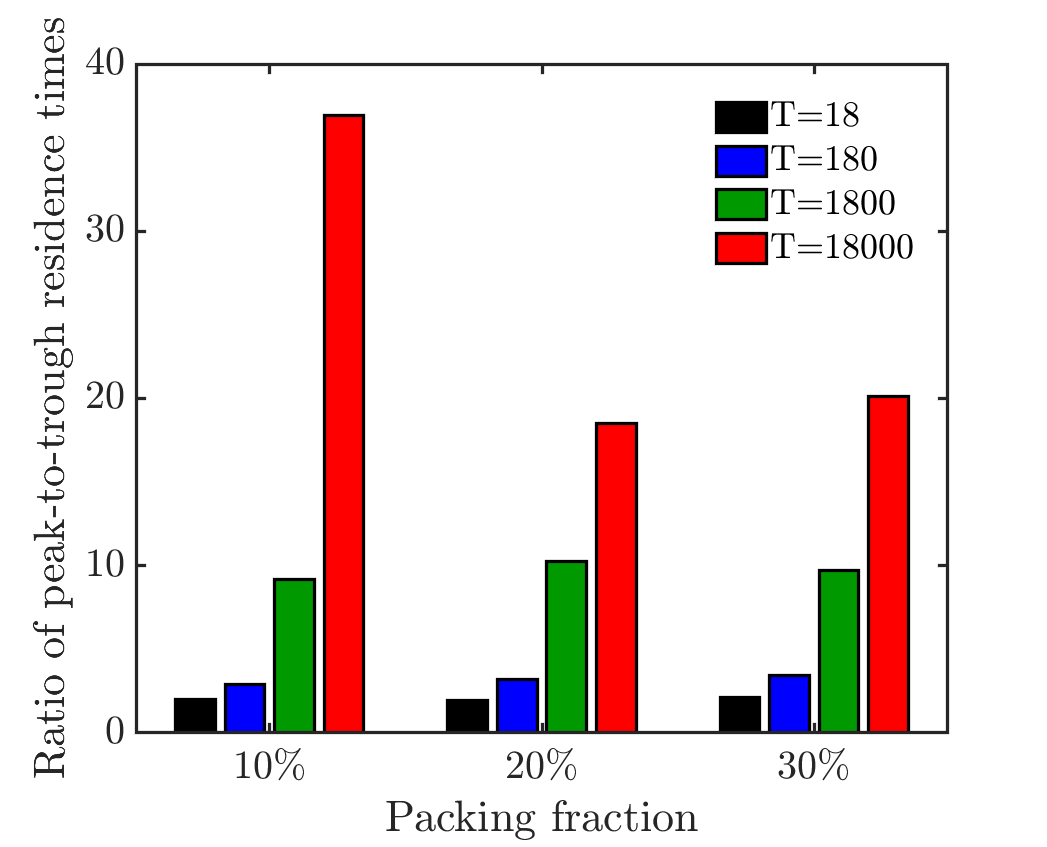}
    \caption{Bar chart describing the ratio of peak-to-trough residence times.}
    \label{barChart}
\end{figure}

 Next we show the results for the residence times in the ordered and disordered states as a function of the oscillation time-period ($T$) and colloid packing fraction. In Fig. \ref{barChart}, the ratio of peak-to-trough durations ($t/T$) are plotted on a bar chart. From this, the results suggest there is a significant imbalance in residence times for systems with larger crosslinker periods. We see that across all three packing fractions, the systems with $T=18000$ spend disproportionately long times in the ordered states and markedly brief times in the disordered states. The system with a packing fraction of $10\%$ and a period of $18000$ has the highest ratio; that is because the system is capable of cycling down to an acutely disordered state that is only maintained for a very brief amount of time. As we decrease $T$, the residence times between the two intervals become smaller as the gap between the separation of the relaxation timescale and crosslinker oscillation timescale is reduced. The smallest period of $18$ leads to systems having comparable times in the peak and trough intervals, but the system did not reach a distinctly well-ordered state and only achieves a disordered state that is weakly connected, with no cluster formation.

\begin{figure*}[h]
  \centering
  \includegraphics[width=2.0\columnwidth]{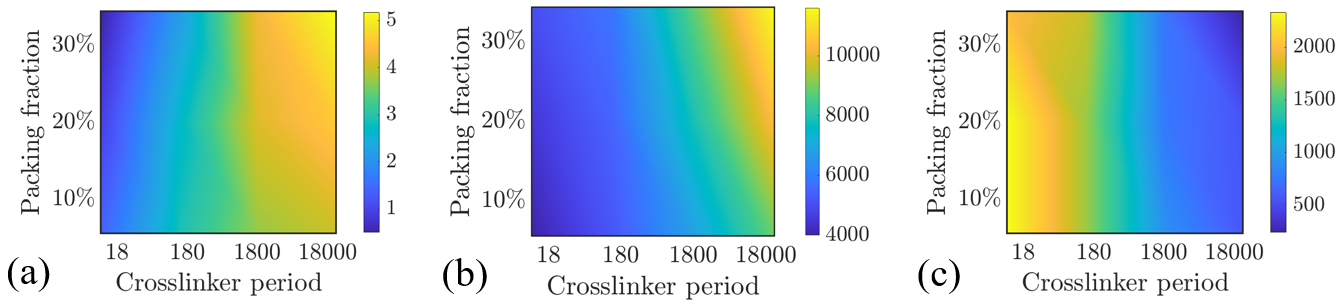}
    \caption{Phase diagrams demonstrating the impact of the packing fraction and crosslinker oscillation period. (a) Phase diagram illustrating the amplitude heights of disordered and ordered states. (b) Phase diagram characterizing the duration of the ordered states. (c) Phase diagram emphasizing the residence times of the relaxed states/at the trough.}
    \label{phaseDiagram}
\end{figure*}

To systematically illustrate how the colloidal packing fraction and crosslinker oscillation period $T$ influence the degree of connectivity in the ordered and disordered states, we show how the amplitude
of the oscillations in the measured connectivity varies with these quantities in Fig. \ref{phaseDiagram}(a)-(c). Figure \ref{phaseDiagram}(a) shows that $T$ controls the differences in amplitude heights. Larger values of $T$ produce systems with pronounced differences in the connectivity of the ordered and disordered states, with the packing fractions having very little effect. The phase diagram shows that the oscillation period of $T=18$ produces the smallest amplitude while the largest oscillation period $T=18000$ yields the greatest amplitude of about $5.5$ bonds per particle.
\par We next show how the residence times in the ordered states are influenced by the packing fraction and $T$ (see Fig. \ref{phaseDiagram}(b)). This is done by measuring the widths of the peaks from the connectivity results, where time is further scaled by $T$. For the smaller crosslinker periods $T=18$ and $180$, all three systems have similar durations spent in the ordered state and the packing fraction has no impact on the lengths of those intervals. As $T$ is increased to $1800$ and $18000$, the systems with the larger packing fractions have longer residence times in the ordered states, with the packing fraction of $30\%$ and $T=18000$ producing the largest residence time. For the smallest packing fraction of $10\%$, the residence times in the ordered states are shortest across all four periods, due to particles being farther apart and having to diffuse more to interact and form bonds. Similar to the patterns of amplitude heights in Fig. \ref{phaseDiagram}(c), the driving factor behind the residence times of the disordered states is the crosslinker period length rather than the packing fraction; we observe that the time spent in the disordered state decreases as the crosslinker time-period is increased and vice-versa.
% \begin{figure*}[t]
%   \centering
%   \includegraphics[width=2.0\columnwidth]{UpdatedFigures/subfigure3_phase_barchart_msd_V3.png}
%     \caption{(a), Phase diagram illustrating the amplitude heights of ordered and ordered periods. (b), Phase diagram characterizing the duration of the ordered periods. (c), Phase diagram emphasizing the residence times of the relaxed states/at the trough. (d), Bar chart describing the ratio of peak-to-trough residence times. (e), The MSD is measured over 4 parts: the first half of the cycle, when particles attach and approach a ordered state, the peak, which is when the system is at its most ordered, the second half of the cycle, when particles began to detach and reach a softer state, and the trough, which is the relaxation period of the colloids.}
%     \label{phaseDiagram_and_MSD}
% \end{figure*}

\section{Conclusion}
We constructed and investigated a colloidal system that can repeatedly undergo order-disorder transitions when rhythmically crosslinked. We explored systems with colloidal packing fractions ranging from $10\%$ to $30\%$ and crosslinker oscillation periods of $18$ to $18000$. We observed sustained oscillations between distinct states with substantially different degrees of connectivity. Larger oscillation periods lead to ordered states with larger connectivity and residence times, but also greater imbalances in residence times between the ordered and disordered states. This is presumably due to several orders of magnitude of separation between the diffusion time of the colloids and the oscillation time-period of the crosslinkers. When these timescales are comparable in magnitude (for example, when $T=18$), the system spends comparable times in the ordered and disordered states but the difference in the degree of order between these two states is very small.

Our results demonstrate that we can achieve distinct states of this colloidal system with pronounced differences in order and material properties when the crosslinker kinetics are governed by the oscillation period. Furthermore, large residence times in the ordered state can be obtained when the crosslinker lifetime depends directly on the oscillation period
 and this oscillation period is much larger than the colloidal diffusion time. These findings provide insights into the rational design of soft materials that can rhythmically transition between states with different material properties on a pre-determined schedule. 

%1052021 lauren
\section*{Conflicts of interest}
There are no conflicts to declare.

\section*{Acknowledgements}
This research  funded by a William M. Keck Foundation Research Grant. 

\balance

%If notes are included in your references you can change the title from 'References' to 'Notes and references' using the following command:
%\renewcommand\refname{Notes and references}

%%%REFERENCES%%%
\bibliography{rsc-articletemplate-softmatterV6} %You need to replace "rsc" on this line with the name of your .bib file
\bibliographystyle{rsc-articletemplate-softmatterV6} %the RSC's .bst file

\end{document}